# Spin helices in GaAs quantum wells: Interplay of electron density, spin diffusion, and spin lifetime

S. Anghel,[1] A. V. Poshakinskiy,[2] K. Schiller,[1] G. Yusa,[3] T. Mano,[4] T. Noda[4] and M. Betz[1]

[1]*Experimentelle Physik 2, Technische Universität Dortmund, Otto-Hahn-Straße 4a, D-44227 Dortmund, Germany*
[2]*Ioffe Institute, St. Petersburg 194021, Russia*
[3]*Department of Physics, Tohoku University, Sendai 980-8578, Japan*
[4]*National Institute for Materials Science, Tsukuba, Ibaraki 305-0047, Japan*
E-mail address: markus.betz@tu-dortmund.de

**Abstract:** To establish a correlation between the spin diffusion, the spin lifetime, and the electron density, we study, employing time-resolved magneto-optical Kerr effect microscopy, the spin polarization evolution in low-dimensional GaAs semiconductors hosting two-dimensional electron gases. It is shown that for the establishment of the longest spin-lifetime, the variation of scattering rate with the electron density is of higher importance than the fulffiling of the persistent spin helix condition when the Rashba α and Dresselhaus β parameters are balanced. More specifically, regardless of the α and β linear dependencies on the electron density, the spin relaxation rate is determined by the spin diffusion coefficient that depends on electron density nonmonotonously. The longest experimental spin-lifetime occurs at an electron density, corresponding to the transition from Boltzmann to Fermi-Dirac statistics, which is several times higher than that when the persistent spin helix is expected. These facts highlight the role the electron density may play when considering applications for spintronic devices.

## I. INTRODUCTION

In low-dimensional semiconductor structures hosting two-dimensional electron gases (2DEGs), the spin-orbit (SO) interaction plays the key role in spin dynamics and relaxation. It is responsible for a broad range of phenomena, including the spin-galvanic effect [1,2], the spin-Hall effect [3], and persistent spin textures [4-6]. The latter ones occur in zinc-blende-type (001)-grown quantum wells (QWs) when the spin-orbit parameters associated with the bulk β (Dresselhaus) [7] and structural α (Rashba) [8] inversion asymmetries are equal in strength [9-11]. Then, the momentum-dependent effective magnetic field $\boldsymbol{B}_{\mathrm{SO}}(\boldsymbol{k})$ associated with the spin-orbit coupling becomes unidirectional and lies in the QW plane. In this specific case, the SU(2) spin rotation symmetry of the system is restored [10] and the dominant Dyakonov–Perel mechanism of spin relaxation is fully suppressed for a particular helical spin distribution. Such a persistent spin helix (PSH), being a long-lived mode, is revealed in the process of spin diffusion at long delay times [10,12,13].

The PSH is usually associated with the condition α=β, which can be achieved either by a precise design of the QW width and modulation doping [14] or by applying the gate voltage that directly affects the Rashba parameter α, ensuring an after-growth fine-tuning. However, a variation of the gate voltage typically also changes the density of electrons in the QW. As a result, also the electron mobility is altered due to a change in the scattering rate, which in turn might affect the dynamics. On the other hand, even in the case of moderate detuning from the condition α=β, the mode with the longest (though finite) lifetime has the helical pattern and is also naturally revealed in the process of spin diffusion [5,12]. The combined impact of changes to the SO couplings and the electron density on the spin dynamics is still unexplored. In this paper, we use time-resolved magneto-optical Kerr effect microscopy (TR-MOKE) to explore the dependence of the spin-orbit parameters, diffusion coefficient and spin relaxation rate on the electron density and search for the conditions to realize the longest lifetime of the helical spin pattern. We demonstrate that the longest spin lifetime is not achieved when fulfilling the condition α=β, but rather in a somewhat detuned case where the spin diffusion coefficient is minimal. This minimum of the spin diffusion coefficient occurs for the electron densities corresponding to the transition from Boltzmann to Fermi-Dirac statistics and can be achieved either by gate voltage tuning or by employing additional optical excitation into the QW barriers.

## II. EXPERIMENTAL DETAILS

The investigated sample is a $n$-doped (001)-oriented 15-nm GaAs quantum well, grown by molecular beam epitaxy and sandwiched between $Al_{0.33}Ga_{0.67}As$ barriers. The QW is patterned in a typical Hall-bar geometry (15 μm wide) with a back-gate and AuGeNi ohmic in-plane contacts. Two Si δ-doping layers are placed above the QW providing a resident electron concentration $n$ in the QW that can be modified by the back-gate voltage $U_{BG}$ [15]. Using photoluminescence spectroscopy, we extract the electron concentration from the Fermi energy of the 2DEG, defined as the energy between the conduction-band minimum and Fermi edge, knowing that $E_\mathrm{F} = \frac{\hbar^2 k^2}{2m^*}$, where $k = \sqrt{2\pi n}$ is the Fermi wavevector of a 2D system and $m^* = 0.064 m_0$ is the effective mass of the electron [16]. We verify that $n$ has an approximately linear dependence on the back-gate voltage in the range of $0\,\text{V} < U_{\text{BG}} < -2.2\,\text{V}$, see Fig. 1(d). The magnitude of the in-plane electric field, useful to measure the electron mobility is set to $E_y = 1.7\,\text{V/cm}$. To extract the Rashba and Dresselhaus parameters an in-plane magnetic field $B_{x(y)} = 220$ mT is used. To create robust electron spins, the sample resides in a compact cold-finger cryostat ensuring a lattice temperature of 3.5 K for all performed measurements.

The time-resolved magneto-optical Kerr microscopy (TR-MOKE) measurements are performed using pulses with a temporal width of ~35 fs derived from a 60 MHz mode-locked Ti:-Sapphire oscillator. Subsequently, they are split into pump and probe paths, which are independently tuned by grating-based pulse shapers [17]. The resulting pulses have a bandwidth of ~0.5 nm and allow for a temporal resolution of ~1 ps. The probe pulses are linearly polarized while the pump pulses are modulated between left ($\sigma^+$) and right ($\sigma^-$) circular polarization by an electro-optic modulator (EOM). Both probe and pump pulses are collinearly focused on the sample surface through a 50× microscope objective. The full width at half-maximum (FWHM) diameter of pump and probe pulses was $w_0 = 3 \pm 0.1$ μm and $1 \pm 0.1$ μm respectively. The reflected pump light is filtered out with a monochromator and the Kerr-rotation of the reflected probe pulse is measured using balanced photodiodes connected to a lock-in amplifier referenced to the EOM frequency. The delay time $t$ between the pump and probe pulses is adjusted by a mechanical delay stage with $t_{\max} = 1.7$ ns. The spatial overlap of the pump with the fixed and centered probe is adjusted through a lateral translation of the input lens of a beam-expanding telescope in the pump path [18,19]. The pump and probe photon energies are chosen based on the spectral response of the 2DEG, see Ref [20].

All measurements are performed with the pump photon energy set to $E_p = 1.57$ eV, which is 40 meV above the optical gap (1.53 eV), and a peak power density of $I_p = 4.7$ MW/cm$^2$. Since the energy separation between the first and second electron levels in the QW is about 52 meV, the pump excites electrons to the first sublevel only. Using 2.6% absorbance of the QW [21] and 30% Fresnel coupling loss, the optically injected electron density is estimated to be $n = 7 \times 10^{11}$ cm$^{-2}$. The probe photon energy is tuned to $E_{\text{pr}} = 1.53$ eV with a pulse peak irradiance of $I_{\text{pr}} = 2.36$ MW/cm$^2$. In some measurments, for an additional optical excitation of carries in the barriers, we use a He-Ne cw laser with an average power density of up to $I_{\text{av}} = 0.56$ kW/cm$^2$ (corresponding to an average power of $P_{\text{av}} = 40$ μW).

## III. RESULTS AND DISCUSSION

Figure 1(a, e) show exemplary temporal evolutions of the spin polarization $S_z$ in the presence of a magnetic ($B_x = 220$ mT) and an in-plane electric field ($E_y = 1.7$ V/cm). Both spin textures are recorded for the back-gate voltage of $U_{\text{BG}} = -1.6$ V, corresponding to an electron density of $n = 1 \times 10^{11}$ cm$^{-2}$. As detailed below, this choice allows for the maximum lifetime of the spin pattern. Without an external electric field, the spin distribution is symmetric with respect to the excitation point. It's center is stationary, but the variance increases with time due to diffusion. An in-plane magnetic field applied to the QW induces a linear shift (tilt of the spatio-temporal pattern) of the spin pattern with time towards $y < 0$ (Fig. 1a). In contrast, an electric field causes the drift of the envelope of the spin distribution towards $y < 0$ (Fig. 1e). More detailed information about the influence of the applied in-plane fields can be found elsewhere [22-24]. In both cases, diffusion and relaxation mechanisms lead to broadening and decay of the spin distribution. The spin distributions along the $y$ coordinate at a given delay time are fitted with the phenomenological equation

$$S_z(y,t) = A\, e^{-\frac{4\ln(2)(y-y_G)^2}{w_y^2}} \cos(2\pi(y - y_c)/\lambda_{\text{SO}}) \quad (1),$$

where $A(t)$ is the amplitude of the spin polarization, $w_y(t)$ is the FWHM of the Gaussian envelope, $\lambda_{\text{so}}(t) = \lambda_0\, w_y(t)^2 / (w_y(t)^2 - w_0^2)$ is the momentary spin precession length, whereas $\lambda_{0,y} = \pi\hbar^2/(m^*|\alpha + \beta|)$ is the precession length of the long-lived spin-helix. $y_G(t) = v_{\text{dr}} t$ describes the drift of the spin distribution envelope in the electric field, whereas the $y_c(t) = v_{\text{ph}} t$ describes the time-varying spatial tilt of the spin pattern induced by the magnetic field. The latter two parameters inherently depend on the applied in-plane fields.

First, we use fits according to Eq. (1) to analyze the spin diffusion in the presence of the magnetic field [Fig. 1(a)]. Figure 1(b) depicts the temporal evolution of the *spin polarization $A\, w_y^2(t)$*. Assuming isotropic spin diffusion, this quantity characterizes the *total* number of spins within the two-dimensional Gaussian distribution. This expression accounts for all initially excited spins, not only those which are oriented along $z$ at the moment of detection and contribute to measured $S_z$ but also those spins, which, due to the spatial precession, lie in the QW plane. This total *spin polarization* allows us to retrieve the spin lifetime by fitting it with a single exponential decay, see the semi-transparent solid line in Fig. 1(b). For the present case, the lifetime is ~3ns – the longest achieved for this structure, see Fig. 3 (d).

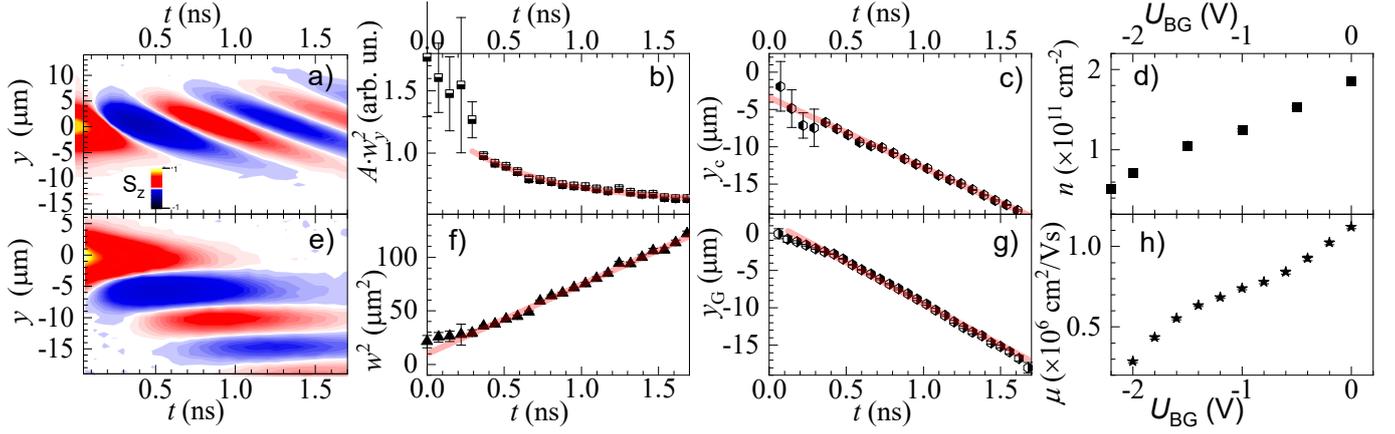

**Fig. 1** Spin polarization micrographs $S_z(x,t)$ measured in **(a)** an in-plane magnetic field of $B_x = 220$ mT and **(e)** an in-plane electric field of $E_y = 1.7$ V/cm. **(b)** The time dependence of the *extracted spin polarization* – $A\,w_y^2(t)$ and its exponential fit (solid line) to extract the spin-lifetime. **(c)** The temporal evolution of spin pattern offset $y_c(t)$ and its linear fit **(f)** The squared FWHM evolution $w_y^2(t)$ together with its linear fit (solid line), used to extract the spin diffusion coefficient $D_s$ according to Eq. (2). **(g)** The spin polarization envelope center $y_G(t)$ evolution and its linear fit, used to determine the phase ($v_{ph}$) and drift velocities ($v_{dr}$). **(d)** and **(h)** represent the back-gate voltage dependence of the electron density and electron mobility, respectively. Side note: panels **(b)**, **(c)** and **(f)** are related to the panel **(a)**, whereas pane **(g)** is related to the panel **(e)**

Figure 1(c) shows the shift of the spin pattern for the case of an applied external magnetic field. The linear dependence of the parameter $y_c$ on the delay time indicates a constant phase velocity according to $y_c(t) = v_{ph}t$. This finding is easily explained by the fact that for the electrons with a certain average momentum $\hbar k_y = m^* v_{ph}$, the external magnetic field $\boldsymbol{B}_x$ is fully compensated by the spin-orbit field $\boldsymbol{B}_{SO}(k_y)$. Hence, the spin precession for these moving electrons is suppressed. This results in the tilted evolution of the spin pattern seen in Fig. 1(a), which effectively depicts the superimposed precession of the spins around the real magnetic field $\boldsymbol{B}_x$ in time and spin-orbit magnetic field $\boldsymbol{B}_{SO}$ in space during diffusion. Taking into consideration the relation $v_{ph} = \frac{\hbar g \mu_B}{2m^*(\alpha+\beta)} B_x$ [4,6], where $g$ is the effective electron $g$-factor and $\mu_B$ is the Bohr magneton, we extract the value of $\alpha + \beta$. Specifically, the $v_{ph}$ dependence on $B_x$ strength shows the expected linear trend (data not shown here), the slope of which returns the strength of the SO-coupling. For this specific case we obtained a value of $\alpha + \beta \approx 4.0$ meV·Å, see Fig. 2(a). This value agrees well with the one extracted from the spatial period $\lambda = 8.8$ μm of spin precession along $y$ direction: $|\alpha + \beta| = \frac{\pi \hbar^2}{m^* \lambda_0} = 4.2$ meV·Å. Similarly, applying the magnetic field in orthogonal direction and measuring the $v_{ph}$ along $x$ (data not shown here) we extract the value of $\alpha - \beta$. This procedure has been repeated for the entire range of the back-gate voltages and both orientations of the external magnetic field. Figure 2(a) and (c) depicts the $(\alpha + \beta)$ and $(\alpha - \beta)$ dependences on the electron densities.

Fig. 1(f) shows the evolution of the width of the Gaussian spin-distribution envelope, which is expected to increase with time due to diffusion according to

$$w_y^2(t) = w_0^2 + 16\ln(2)\,D_s t \qquad (2),$$

where $D_s$ is the spin diffusion coefficient and $w_0$ is the initial FWHM, defined by the laser spot. A linear fit to Eq. (2) allows us to retrieve the spin diffusion coefficient $D_s$. We note that the data presented in Fig. 1(f) shows the temporal evolution of the Gaussian envelope in the presence of an in-plane magnetic field rather than in the presence of the in-plane electric field. This was done because it is well documented that an applied in-plane electric field leads to the electron gas heating and modification of the local electron mobility which alters the diffusion picture [22,24].

Figure 1(g) shows the drift of the spin polarization envelope center due to the applied in-plane electric field – the fitting results of the $S_z$ in Fig. 1(e). The data depict a linear trend, which enables the drift velocity extraction according to $y_G(t) = v_{dr}t$. The latter one combined with the known magnitude of the electric field $E_y = 1.7$ V/cm allows us, in turn, to calculate the electron mobility $\mu$. Figure 1(h) shows the dependence of $\mu$ on the back-gate voltage in the same range of voltages as the electron densities $n$ dependence [see Fig. 1(d)], namely $0\,\text{V} < U_{BG} < -2.2\,\text{V}$. Given the fact that the electron density $n$ exhibits a nearly linear dependence on the back-gate voltage, all the subsequent data will be presented in dependence on electron concentration rather than back-gate voltage.

Figure 2(a) shows the dependence of the $\alpha + \beta$ parameter on the electron density; the data were extracted according to the procedure described above. The only parameter that directly depends on $n$ is the Dresselhaus parameter $\beta = \beta_1 - \beta_3$, via $\beta_3 = \gamma_D n\pi/2$ that is the $\mathbf{k}$-cubic contribution present in QWs alongside with the $\mathbf{k}$-linear contribution $\beta_1$, which in turn does not depend on $n$ [6,25]. Here, $\gamma_D = (9 - 11)$ meVÅ$^3$ is the Dresselhaus coefficient for GaAs QWs. However, concomitantly, the back-gate voltage changes the $E_z$ electric field component perpendicular to the QW plane. The latter one is responsible for modification of the Rashba parameter $\alpha = \gamma_R E_z$, where $\gamma_R$ is the Rashba coefficient. While the electron density is readily determined from the photoluminescence measurements, the estimation of the electric field $E_z$ is a much more difficult task. Consequently, the data presented in Fig. 2 are shown in dependence on the electron concentration only, although both $n$ and $E_z$ contributions are present. Figure 2(a) reveals a linear trend for the $\alpha + \beta$ parameter, with 1.5 times decrease when the electron concentration is changed in the given range.

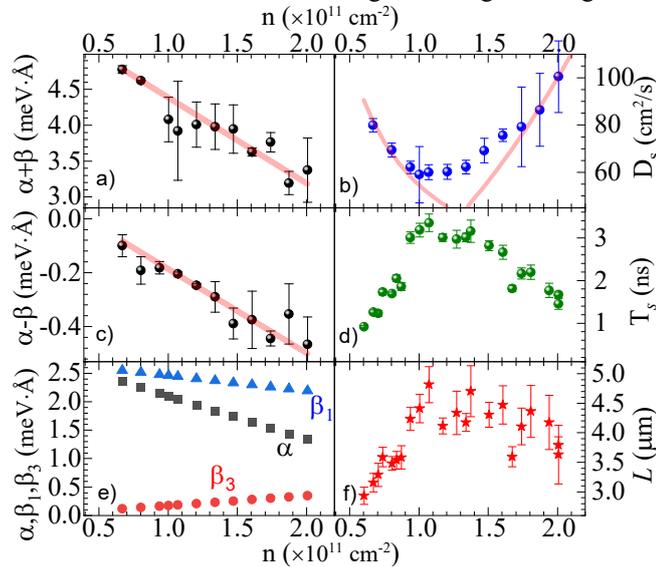

**Fig. 2** Experimentally measured **(a)** spin-orbit parameter $\alpha + \beta$, **(b)** spin diffusion coefficient $D_s$, **(c)** spin-orbit parameter $\alpha - \beta$ and **(d)** spin-lifetime $T_s$ as a function of the electron density. The solid lines in **(a)** and **(c)** are linear fits to the data, whereas those in **(b)** are fits for low- and high-density regions according to Eq. (8) and (9), respectively. **(e)** The individual spin-orbit parameters: experimentally extracted $\alpha$ and $\beta_1$ and the calculated $\beta_3$. **(f)** The spin-diffusion length calculated accordingly to $L_S = \sqrt{T_s D_s}$.

The $\alpha - \beta$ dependence shows a similar linear behavior, although this time its absolute value increases with increasing $n$, see Fig. 2(c). The spin-orbit parameters $\alpha, \beta_1$ and $\beta_3$ are shown individually in the Fig. 2(e) – the latter parameter has been calculated as described in the above discussion, taking into consideration the electron density changes, whereas the former two parameters were extracted from the combination of the linear fits of the experimental data presented in Fig. 2 (a) and (c). As expected, the $\beta_1$ does not vary to much with the electron concentration, in marked contrast to $\alpha$ and $\beta_3$ parameters, which do strongly depend on $n$. For a more in-depth discussion about the influence of the transversal electric field and the electron concentration on the spin-orbit parameters the readers are redirected elsewehere[14,26].

However, more importantly, the data presented in Fig. 2(c) allows us to determine the electron density where the PSH regime is established, i.e., where $\alpha = \beta$. By linearly fitting the data for $\alpha - \beta$, the latter condition is reached for $n = 0.3 \times 10^{11}$ cm$^{-2}$. In this case, the lifetime of the longest-lived helical spin mode, determined by [4]

$$T_{SH}^{-1} \approx 2D_s \frac{m^{*2}}{\hbar^4}[3\beta_3^2 + (\alpha - \beta)^2] \quad (3a),$$

is expected to be maximal. We note that Eq. (3a) contains only the spin-orbit parameters $\alpha$ and $\beta$ and the spin-diffusion coefficient $D_s$.

The experimentally extracted $D_s$ and spin-lifetime $T_s$ are shown in Fig. 2(b and d), respectively. Both graphs show nonmonotonic dependences on the electron concentration, with $D_s$ reaching the minimum for $n \approx 1.2 \times 10^{11}$ cm$^{-2}$; at the same density the spin-lifetime $T_s$ reaches a maximum. This is in contradiction to a naïve picture where the maximum spin-lifetime is expected to occur for the PSH point, $\alpha = \beta$, realized at the electron density $n = 0.3 \times 10^{11}$ cm$^{-2}$ (see Fig. 2c) which is a few times smaller than the maximum position of the experimental $T_s$ curve. To even more emphasize this fact, the Fig. 2(f) shows the spin-diffusion length $L_S = \sqrt{T_s D_s}$ dependence on the same electron concentration. A naïve prediction would be that $L_S$ should be determined by spin-orbit constants only, thus vary slowly with electron concentration. That is approximately satisfied in the high electron concentration regime where $L_S$ is roughly constant. At small concentrations, as PSH condition is approached, $L_S$ does not increase but is suppressed instead. The turning point happens again, as in the case of spin-lifetime and spin diffusion coefficient, at the $n = 0.3 \times 10^{11}$ cm$^{-2}$, highlighting this surprising and interesting result.

This discrepancy with the idealized theoretical model of PSH becomes even more evident from Fig. 3a, where the three different spin lifetimes are compared (note the logarithmic scale). The green circles represent the experimental results (same as in Fig. 2c) while the red triangles are the spin helix lifetime $T_{SH}$ calculated according to Eq. (3a), taking into consideration the experimentally extracted values of $\alpha, \beta_1, \beta_3$ and $D_s$ from Fig. 2. As expected, the spin helix lifetime is increased when the PSH condition ($n = 0.3 \times 10^{11}$ cm$^{-2}$, out of the plot scale) is approached. It is worth mentioning that the obtained time $T_{SH}$ is longer by more than an order of magnitude than the experimentally measured lifetime. The lifetime $T_{SH}$ describes the decay of the most robust mode that dominates in the spin-diffusion picture only at extremely large times $t \gtrsim T_{SH}$, a condition hardly accessible in the experiment. The decay of the spin distribution at smaller times, measured in the experiment, is determined by other spin modes with smaller lifetimes. For comparison, we plot (stars in Fig. 3a) the calculated lifetime of the uniform spin distribution $S_z$ [27]

$$T_{s0}^{-1} \approx 4D_s \frac{m^{*2}}{\hbar^4}(2\beta_3^2 + (\alpha-\beta)^2 + (\alpha+\beta)^2) \quad (3b),$$

which is likely to determine spin decay shortly after photoexcitation ($t \lesssim T_{s0}$), when the spin helix pattern has not yet formed. Indeed, a simple approximate equation describing spin dynamics at all times derived in Ref. [28] yields that the logarithmic time-derivative of the total spin $A w^2$ at zero time should equal $T_{s0}^{-1}$. The obtained values of $T_{s0}$ are an order of magnitude shorter than the experimentally measured lifetime. We note that, in contrast to $T_{SH}$, the time $T_{s0}$ has no peculiarity at the PSH condition $\alpha = \beta$, cf. Eqs. (3a) and (3b), and even slightly decreases when the PSH case is approached.

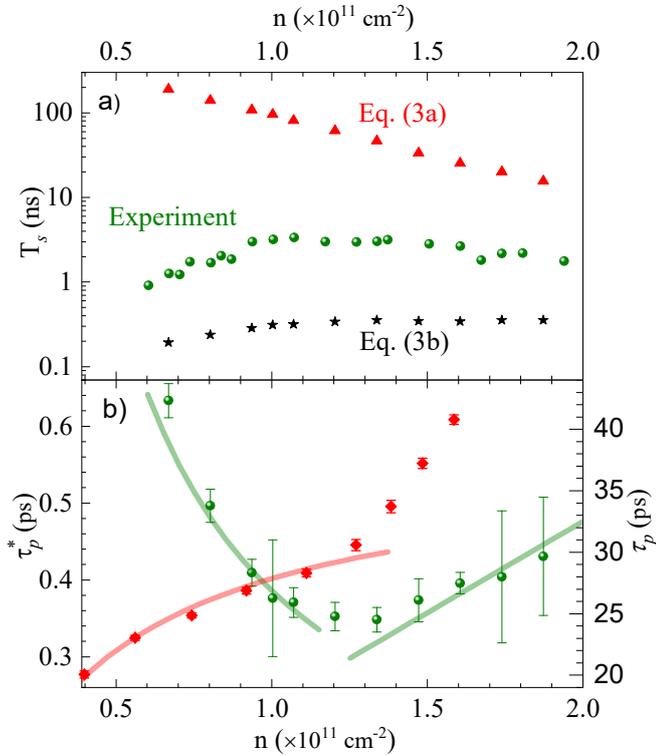

**Fig. 3 (a)** Experimental spin lifetime dependence together with the calculated ones according to Eqs. (3a) and (3b). **(b)** Momentum scattering time $\tau_p$ (diamonds) and scattering time accounting electron-electron collisions $\tau_p^*$ (dots) in dependence of the electron density $n$ with the fits according to Eq. (5) and Eq. (6), respectively. The $\tau_p^*$ fitting is shown as two asymptotes, $\tau_p^* \sim 1/n$ and $\tau_p^* \sim n$, corresponding to the regions of low and high electron densities.

For the experiment, the intermediate times $T_{s0} \ll t \ll T_{SH}$ are relevant. The experimentally measured lifetime dependence lies in-between the theoretical curves calculated for the two limiting cases according to Eqs. (3a) and (3b). We also note that the experimental lifetime dependence shares the features of both $T_{SH}$ and $T_{s0}$: it grows with electron density at low densities similar to the latter and decays at high densities similar to the former. Interestingly, the maximum value of the experimentally measured lifetime $T_s$ is achieved not at the PSH condition, as it would be for $T_{SH}$, but rather at the point close to the spin diffusion coefficient minimum, see Fig. 2(b) and (d). This highlights the importance of considering the spin diffusion coefficient dependence on the electron density when designing spintronic devices.

Below, we explain the non-monotonous dependence of $D_s$ (and consequently that of $T_s$) on electron density by the impact of electron – electron scattering and considering the transition of the electron gas statistics from the Boltzmann case (at low electron densities) to the Fermi-Dirac case (at higher electron densities). In principle, both the electron density and temperature after optical excitation are spatially and temporally dependent, which leasds to the variation of all spin transport parameters – spin diffusion coefficient, spin mobility, the spin-orbit interaction strength. Previously, we have shown that the spatial variation of spin mobility leads to anisotropic spread of electron distribution [22], while the spatial variation of spin-orbit parameters results in the *s*-shaped stripes in the spatio-temporal maps of spin diffusion [29]. In the present experiments, no such effects are revealed. Therefore, for the sake of simplicity, we neglect the spatial variation of the electron gas parameters and describe it by some effective spatially homogeneous electron density and temperature.

The spin diffusion coefficient in both Fermi-Dirac and Bolzmann cases is determined by the average electron velocity $\langle v^2 \rangle$ and the scattering time $\tau_p^*$:

$$D_s = \frac{1}{2}\tau_p^*\langle v^2\rangle = \frac{\xi \tau_p^*}{m^*(1-e^{-\xi/k_B T})} \quad (4),$$

where $\xi$ is the chemical potential of electrons and $T$ is the effective electron temperature. The scattering time $\tau_p^*$ can be presented in the form

$$\frac{1}{\tau_p^*} = \frac{1}{\tau_p} + \frac{1}{\tau_{ee}} \quad (5),$$

where $\tau_p$ is electron momentum scattering time due to collisions of electrons with impurities, phonons, and photogenerated holes, and $\tau_{ee}$ is the time of electron-electron scattering which affects spin diffusion due to the spin Coulomb drag effect [30,31]. Note that $\tau_p$ can be measured directly in the spin drift experiment (Fig. 1(h)). The momentum scattering time extracted from the measured mobility using the identity $\mu = e\tau_p/m^*$ is shown in Fig. 3(b) as red symbols. We fit the experimental points with the dependence

$$\frac{1}{\tau_p} = \frac{1}{\tau_p^{(0)}} + \frac{\alpha}{n}, \quad (6),$$

see solid red line in Fig. 3(b). The first term in Eq. (6) describes the energy-independent contribution to the momentum scattering rate originating from, e.g., the scattering by short-range impurities or phonons, and the second term $\sim 1/n$ might originate from the Coulomb impurities. At high electron densities $n > 1.2 \cdot 10^{11}$ cm$^{-2}$, $\tau_p$ abruptly grows, which might be related to screening of the impurity potential.

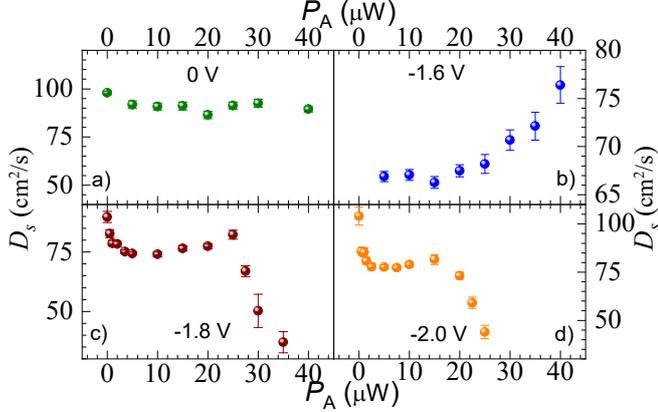

**Fig. 4** Spin diffusion coefficient $D_s$ in dependence of additional optical excitation with a He-Ne CW laser for four different cases of back-gate voltages – **(a)** $U_{BG} = 0$V, **(b)** $U_{BG} = -1.6$V, **(c)** $U_{BG} = -1.8$V and **(d)** $U_{BG} = -2.0$V.

The electron-electron scattering time has distinct forms for the cases of Boltzmann and Fermi-Dirac statistics [32]:

$$\frac{1}{\tau_{ee}} = \begin{cases} \dfrac{c_1 e^4 n}{\hbar \varepsilon^2 k_B T}, & \xi \ll k_B T, \\ \dfrac{c_2 (k_B T)^2}{\hbar \xi}, & \xi \gg k_B T \end{cases} \quad (7),$$

where $\varepsilon$ is the dielectric permittivity, $c_{1,2}$ are slowly varying dimensionless functions that are determined by structure parameters. The calculation yields (see green dots in Fig. 3(b) and the discussion below) that in both regimes the electron-electron scattering time is at least an order of magnitude shorter than $\tau_p$. Therefore, the dominant contribution to the right-hand side of Eq. (5) is given by the electron-electron collisions. Then, the spin diffusion coefficient for Boltzmann and Fermi-Dirac statistics assume the form

$$D_s \sim \begin{cases} \dfrac{\hbar \varepsilon^2 (k_B T)^2}{c_1 e^4 m^* n}, & n \ll \dfrac{m^* k_B T}{\pi \hbar^2} \quad (8) \\ \dfrac{\pi^2 \hbar^5 n^2}{c_2 m^{*3} (k_B T)^2}, & n \gg \dfrac{m^* k_B T}{\pi \hbar^2} \quad (9). \end{cases}$$

It follows from Eq. (8) and (9) that at low electron densities the spin diffusion coefficient decreases with the density as $D_s \sim 1/n$, while at high densities it grows as $D_s \sim n^2$, which explains the origin of the observed non-monotonous dependence.

We now fit the spin diffusion coefficient to Eqs. (8)-(9) taking $c_1 = \pi^2/2$ and $c_2 = (\pi/4) \ln(\xi/k_B T)$ as a simplest rough estimation [32], see the solid curve in Fig. 2(b).

In the region of small electron densities, we use Eq. (8) and obtain the effective electron temperature $T \approx 110$ K. Such high temperature is explained by the large number of hot electrons generated by optical pump and probe pulses [33,34]. In the region of high densities, using Eq. (9), we obtain the effective electron temperature $T \approx 40$ K, which is a few times smaller than the effective temperature for low electron densities. The decrease of the electron heating efficiency agrees well with the fact that, for the same generation rate of the hot photoelectrons, a larger initial electron density leads to smaller average electron energy. We shall note that for the lowest considered density we have $\xi/k_B T = \ln[\exp(\pi \hbar^2 n / k T m^*) - 1] \approx -1$ while for highest density $\xi/k_B T \approx 2$. This indicates, that there are deviations of the electron statistics from the exact Boltzmann and Fermi-Dirac cases, which might explain the small difference between the theoretical and experimental dependencies of $D_s$.

In Fig. 3(b), we plot the electron-density dependence of the scattering time $\tau_p^*$, extracted from the $D_s$ dependence in Fig. 2(b) using Eq. (4). As stated above, the time $\tau_p^*$ appears to be much shorter than the momentum scattering time $\tau_p$ for the entire range of the electron density (note the time scales in Fig. 3(b)). Similar to the $D_s$ dependence, $\tau_p^*$ depends on the electron density non-monotonously, marking the transition between Boltzmann and Fermi-Dirac statistics. Solid semitransparent lines show the asymptotes $\tau_p^* \sim 1/n$ and $\tau_p^* \sim n$ corresponding to the regions of low and high electron densities [see Eq. (7)].

To further corroborate the influence of the electron density on spin-lifetime, we also study the $D_s$ dependence on additional optical pumping of carries done with a He-Ne cw laser, i.e. in the barriers of the QW. We checked that the correlation between the experimentally measured spin helix decay rate and the spin diffusion coefficient discussed above also holds in this case. Given the fact that the exact numbers of the layers and thus the thickness of the material involved in the absorption process of the additional optical excitation is difficult to estimate, we plot in Fig. 4 the dependence of spin diffusion coefficient on the optical pumping power $P_A$ rather than the electron density. The panels correspond to four different values of the back-gate voltage. Figure 4(a) shows the case of $U_{BG} = 0$ V, corresponding to an intrinsic electron concentration of $n = 2.0 \times 10^{11}$ cm$^{-2}$ and an intrinsic $D_s = 100$ cm$^2$/s (see Fig. 2 (b)). As can be seen from Fig. 4(a), $D_s$ does not markedly depend on additional optical pumping. This indicates that with no applied back-gate voltage, the electrons created in the barriers probably do not relax into the QW and thus have no influence on the processes taking place in the QW.

Fig. 4(b) depicts the case of $U_{BG} = -1.6$V, namely the situation corresponding to the longest spin-lifetime and smallest spin diffusion coefficient. For this back-gate voltage, the electron density is $n \approx 1.1 \times 10^{11}$ cm$^{-2}$ and the spin diffusion coefficient reaches its minimum value of $D_s \approx 65$ cm$^2$/s (see Fig. 2(b)). For such negative back-gate voltages, the electrons created by the additional optical pumping will relax into the QW. An increase of the average pumping power means an increase of the electrons concentration, thus it is expected that $D_s$ should increase as well, as if moving alongside the right branch of the parabola-like dependence in Fig. 2(b). Indeed, Fig. 4 (b) shows exactly this behavior; the $D_s$ increases from $D_s \approx 65 \text{ cm}^2/\text{s}$ to a value of $D_s \approx 80 \text{ cm}^2/\text{s}$. Moreover, the latter value can be used to tentatively estimate the total electron concentration (the initial one induced by the back-gate voltage and the additional one created by the optical pumping): the value of $D_s \approx 80 \text{ cm}^2/\text{s}$ corresponds to an electron concentration of $n = 1.8 \times 10^{11} \text{ cm}^{-2}$ in Fig. 2(b). Thus, for this specific voltage and pump power range the electron concentration is increased roughly by 60%.

Now, we turn towards the low-density region, i.e., the left side of the $D_s$ minimum in Fig. 2(b), which is realized for higher values of the back-gate voltages. In this case, an increase of the electron concentration, via optical pumping intro barriers, should force the $D_s$ to follow the same path as in Fig. 2(b), namely $D_s$ should decrease until it reaches a minimum and then increase again in a parabola-like shaped dependence. Fig. 4 (c and d) show the $D_s$ dependence on optical pumping for two different back-gate voltages $U_{\text{BG}} = -1.8\text{V}$ and $U_{\text{BG}} = -2.0\text{V}$, respectively. At low pumping, both experimental dependencies follow the expectations, i.e. there is a sharp decrease of $D_s$ up to $P_A \approx 5 \text{ μW}$ where it reaches a minimum, followed by an increase up to $P_A \approx 20 \text{ μW}$. However, increasing the optical pumping power further, above the $P_A \approx 20 \text{ μW}$, leads to a sharp decrease of $D_s$. This is in stark contrast with Fig. 4(b) as well as with what would be expected from the dependence of Fig. 2(b). To tentatively explain this behavior, we should recall that the back-gate voltage in the latter two cases is higher compared to the case presented in Fig. 4(b). This higher electric field perpendicular to the QW plane translates to a higher number of hot-electrons being dragged by this field into the QW from the barriers where they are generated. As was mentioned before, the hot electrons, having huge excess energy, can rise the effective temperature of the 2D electron gas [33]. This boosts the electron-electron interaction and decreases the scattering time $\tau_p^*$ which in turn decreases $D_s$. The higher the back-gate voltage across the QW, more hot electrons are dropping into it and thus the scattering time goes down faster. The fact that the kink in Fig. 4(d) happens at lower pump power $P_A \approx 16 \text{ μW}$ compared to that in Fig. 4(c) $P_A \approx 25 \text{ μW}$ confirms the above picture.

Another way to corroborate the influence of the electron density on spin-lifetime may be to rise the lattice temperature. Unfortunately, in this case, higher lattice temperatures mean that the electron-phonons scattering starts to play an important role for all spin transport parameters (the spin diffusion coefficient, spin mobility, the spin-lifetime, etc.) and this will hinder the pure influence of the electron concentration on spin-lifetime.

## IV. CONCLUSIONS

In conclusion, we have analyzed how the properties of spin helices in a 2DEG depend on the electron density, which can be tuned either by applying a gate voltage or by additional optical pumping. The change of the electron density leads to the variation of the spin-orbit parameters, which determine the period of the spin helix and the spin diffusion coefficient that governs the spatial spread of the spin helix. The decay of the spin distribution at very long delay times is expected to be determined by the lifetime of the long-lived spin mode that reaches a maximum for a particular electron density when Rashba and Dresselhaus parameters are balanced $\alpha = \beta$. However, the actual spin helix decay, accessible in the experiment, is much faster and depends on the spin diffusion coefficient. We have found that the latter is governed by electron-electron scattering and has a strongly non-monotonous dependence on electron density, explained by the transition from Boltzmann to Fermi-Dirac statistics. In the transition region, the value of the spin diffusion coefficient reaches a minimum while the spin helix lifetime is at the maximum. These findings are useful to optimize the parameters of future spintronic devices.

## ACKNOWLEDGEMENTS

We acknowledge fruitful discussions with S.A. Tarasenko and M. Matsuura. This work was supported by the Deutsche Forschungsgemeinschaft (DFG) in the framework of the ICRC – TRR 160, project B3. Theoretical considerations were supported by the Russian Science Foundation grant No. 21-72-10035. AVP also acknowledges support from the the Russian Foundation for Basic Research (project No. 19-52-12038) and the RF Grant No. MK-4191.2021.1.2. We also acknowledge the support from the Grant-in-Aid for Scientific Research (No. 17H01037, 19H05603, 21H05188, 21F21016) from the Ministry of Education, Culture, Sports, Science, and Technology (MEXT), Japan.